\documentclass[prb,twocolumn]{revtex4-1}

\usepackage{graphicx}
\usepackage{multirow}
\usepackage{dcolumn}
\usepackage{bm}
\usepackage[normalem]{ulem}
\usepackage{amsmath}
\usepackage{amsfonts}
\usepackage{amssymb}
\usepackage{color}
\usepackage{colordvi}
\usepackage{dcolumn}

\renewcommand{\L}{\mathrm{L}}
\newcommand{\X}{\mathrm{X}}

\begin{document}

\title{Cascade of phase transitions in the vicinity of a quantum critical
point}
\date{\today }
\author{H. Meier$^{1,2}$, C. P\'{e}pin$^{3}$, M. Einenkel$^{2}$, and K. B. Efetov$^{2,3,\Diamond}$}
\affiliation{
$^1$Department of Physics, Yale University, New Haven, Connecticut 06520, USA\\
$^2$Institut f\"ur Theoretische Physik III, Ruhr-Universit\"at Bochum,
44780 Bochum, Germany\\
$^3$IPhT, CEA-Saclay, L'Orme des Merisiers, 91191 Gif-sur-Yvette, France\\
$^\Diamond$ Corresponding author, E-mail : efetov@tp3.rub.de.}

\begin{abstract}
We study the timely issue of charge order checkerboard patterns observed in a variety of cuprate superconductors. We suggest a minimal model in which strong quantum fluctuations
in the vicinity of a single antiferromagnetic quantum critical point generate the complexity seen in the phase diagram of cuprates superconductors and, in particular, the
evidenced charge order. The Fermi surface is found to fractionalize into hotspots and antinodal regions, where physically different gaps are formed. In the phase diagram, this is
reflected by three transition temperatures for the formation of pseudogap, charge density wave, and superconductivity (or quadrupole density wave if a sufficiently strong magnetic field is applied). The charge density wave is characterized by modulations along the bonds of the CuO lattice with wave vectors connecting points of the Fermi surface in the antinodal regions. These features, previously observed experimentally, are so far unique to the quantum critical point in two spatial dimensions and shed a new light on the interplay between strongly fluctuating critical modes and conduction electrons in high-temperature superconductors.
\end{abstract}

\maketitle

\section{Introduction}

High-temperature (high-$T_{c}$) cuprate super\-conductors \cite{mueller,nagaosa,np} rank
among the most complex materials ever discovered. Despite the rich diversity
within the cuprate family, all compounds share common features such as
the antiferromagnetic Mott insulator phase at zero or small doping.
Magnetic fluctuations are ubiquitously present in all compounds of
the cuprate family. Upon hole-doping of the copper-oxide planes, they become
superconductors at unusually high transition temperatures~$T_c$. Ultimately, at intermediate doping,
they exhibit the enigmatic pseudo-gap phase characterized by a gap observed in
transport and thermodynamics up to a temperature~$T^*>T_c$.

In the last years, incommensurate charge modulations have been reported in many
of the families' compounds. These modulations form a checkerboard pattern and possibly also break nematicity.
\cite{wise,davis,yazdani,julien,ghiringhelli,chang,achkar,leboeuf,blackburn}
Complimentary to each other, these experiments demonstrate
that this order is different from stripe spin-charge modulations predicted earlier \cite{zaanen,machida},
observed in La-compounds \cite{tranquada}, and discussed in
numerous publications (see, e.g., Refs.~\onlinecite{emery,kivelson}), as well as from the
$d$-wave order proposed in Ref.~\onlinecite{chakravarty}.

Among the simplest
properties common to all the cuprate compounds is the presence of strong
antiferromagnetic fluctuations due to the proximity of a doping-driven quantum phase
transition between an antiferromagnetic and normal metal phase.
Approaching the complexity of the cuprates
from the perspective of this \emph{universal} singularity, we provide an extensive
study of a single antiferromagnetic two-dimensional quantum critical point (QCP).
Proximity to quantum phase transitions \cite{sachdev,sachkeim} is generally believed important
to explain the intriguing behavior of high-$T_c$
cuprates \cite{nagaosa,np,mm}, heavy fermions \cite{lrvw}, or doped
ferromagnets \cite{mac}.

Our study unveils that this QCP triggers a cascade of phase transitions with
symmetries different from those of the parent transition. These
phases include $d$-wave superconductivity, a checkerboard structure of quadrupole density wave (QDW),
a charge density wave (CDW) with another checkerboard
structure turned by $45^\circ$ with respect to the former, and the ``pseudogap state''
which lacks any long range order. The additional charge order (CDW) arises due to
interaction of electrons with superconducting fluctuations in situations when
superconductivity itself is destroyed. To the best of our knowledge, 
formation of CDW due to superconducting fluctuations 
has not been considered previously. 
The complexity of the phase diagram is
recovered out of a single original QCP using a low energy effective theory
describing interaction between low energy fermions and paramagnons, which represent the quantum fluctuations of the antiferromagnetic order parameter.
This unexpected
result enriches the conventional picture \cite{sachdev,sachkeim} of a single QCP and
may provide new insights into the pseudogap phase of hole-doped cuprates.

\section{Physical picture}

\label{sec:picture}

Before delving into details of the microscopic derivation, let us first develop the physical picture and phenomenology.
In Sec.~\ref{sec:micro} we provide
a microscopic study to back up the physical picture, and finally, in Sec.~\ref{sec:cuprates} we address the question how our results
may help to understand physical phenomena observed in the high-$T_c$ cuprates.

\subsection{Spin-fermion model and pseudogap state}

We adopt the two-dimensional spin-fermion model \cite{ac} for the antiferromagnetic QCP
as the ``minimal model'' in which we seek to understand the diversity of the non-magnetic phases.
As has been known for a while \cite{ac,acs}, this model features a superconducting instability of the normal metal state.
More recently, linearizing the quasiparticle spectrum near so-called ``hotspots'', Metlitski and Sachdev pointed
out \cite{ms2} an $\mathrm{SU}(2)$ particle-hole symmetry of the effective Lagrangian that might
lead to another instability toward a ``bond
order'' state. About two years later, it has been noticed \cite{emp} that, in fact, a state with a complex order parameter
comprising both superconductivity and
an unusual charge order forms below a certain $T^*$. These phenomena significantly expand the earlier effective picture \cite{hertz} of free but Landau-damped
paramagnons.

\begin{figure}[t]
\centerline{\includegraphics[width=\linewidth]{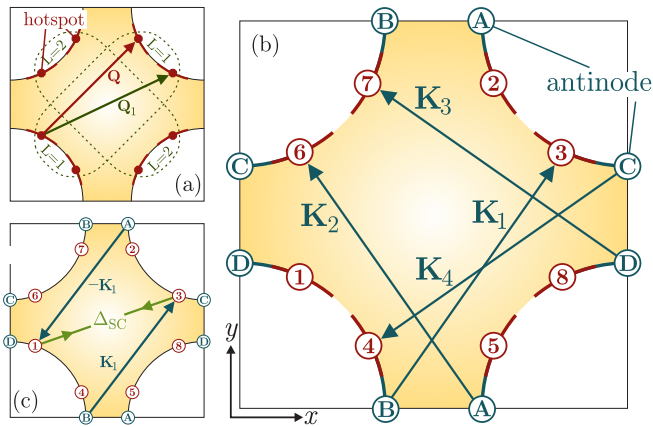}}
\caption{(a) Brillouin zone and Fermi surface. Quantum
critical paramagnons single out eight hotspots that we organize in two
quartets ($\mathrm{L} =1$ and $\mathrm{L} =2$). (b)
Extended model of hotspot (red) and antinodal states (blue). Non-singular
paramagnons with wave vectors~$\mathbf{K}_{1},\ldots,\mathbf{K}_{4}$
mediate the interaction between hotspot and antinodal states.
(c) Cooper pair generation at antinodes~$\mathbf{A}$ and~$\mathbf{B}$.}
\label{fig01}
\end{figure}

In the model considered, spin-$\tfrac{1}{2}$ fermion quasiparticles~$\psi=(\psi_\uparrow,\psi_\downarrow)$, which occupy states close 
to the Fermi surface shown in Fig.~\ref{fig01}(a),
couple to paramagnons~$\boldsymbol{\phi}=(\phi^x,\phi^y,\phi^z)$ with propagator
\begin{align}
\big\langle\phi^\alpha_{\omega,\mathbf{q}}\phi^\beta_{-\omega,-\mathbf{q}} \big\rangle
 &= 
\frac{\delta_{\alpha\beta}}{c^{-2}\omega^2+(\mathbf{q}-\mathbf{Q})^2+\xi_{\mathrm{AF}}^{-2}}
\ ,  \label{a01}
\end{align}
where~$c$ is the velocity of paramagnon excitations. At the QCP, the length $\xi_{\mathrm{AF}}$
diverges so that the paramagnon propagator becomes
singular at the antiferromagnetic ordering wave vector~$\mathbf{Q}=(\pm \pi
/a,\pm \pi /a)$, where $a$ is the lattice constant of the Cu layer. Quasiparticles
emitting or absorbing such singular paramagnons exist
only in the vicinity of eight hotspots, see Fig.~\ref{fig01}. In a first approximation, we  thus
focus on these hotspots.

The energy scale~$\Gamma\sim\lambda^2$, where~$\lambda$ is fermion--paramagnon coupling constant,
determines a temperature~$T^*\sim 0.1\Gamma$, below which
a complex order with competing charge and
$d$-wave superconducting suborders shows up \cite{emp} and completely changes major properties of the system. 
The order parameter in this regime can be represented in the form $b_0 u$, where
$b_0\sim\Gamma$ is an amplitude and $u$ an unitary matrix in particle-hole space,
\begin{align}
u= \left(
\begin{array}{cc}
\Delta _{\mathrm{QDW}} & \Delta _{\mathrm{SC}} \\
-\Delta _{\mathrm{SC}}^* & \Delta _{\mathrm{QDW}}^*
\end{array}
\right) \ .  \label{k0}
\end{align}
In this matrix, $\Delta_{\mathrm{QDW}}$ and $\Delta_{\mathrm{SC}}$ are complex amplitudes for
charge order and superconductivity, respectively. Unitarity imposes
$|\Delta_{\mathrm{QDW}}|^{2}+|\Delta_{\mathrm{SC}}|^{2}=1$. In fact,
there are two independent order
parameters of the form of Eq.~(\ref{k0}), one for each of the two quartets
of hotspots, Fig. \ref{fig01}(a), which in the
``hotspot-only'' approximation are effectively decoupled.

The charge order competing with superconductivity is characterized by a
quadrupole moment spatially modulated with wave vectors~$\mathbf{Q}_1$ and~$\mathbf{Q}_2$, see Fig.~\ref{fig02}.
These wave vectors connect hotspots opposite to each other
with respect to the center of the Brillouin
zone but are equivalently represented as in the inset of Fig.~\ref{fig02}. The
resulting checkerboard structure of this \emph{quadrupole-density wave} (QDW) is shown in Fig.~\ref{fig02}(a).
The QDW (or, equivalently, ``bond order'') instability for wave vectors~$\mathbf{Q}_{1,2}$ has
been recently confirmed in an unrestricted Hartree-Fock study. \cite{sachdev13a}

The matrix order parameter $u$, Eq.~(\ref{k0}), obtained from mean-field
equations at temperatures $T<T^*$ is highly degenerate. \cite{emp} At low enough temperatures,
this degeneracy is lifted by curvature and magnetic field effects, the former
favoring superconductivity, the latter QDW. \cite{mepe} At high enough temperatures (but still below $T^*$)
thermal fluctuations restore the degeneracy and thus establish a \emph{pseudogap phase} without a specific long-range order.
The effective $\mathrm{O}(4)$ non-linear $\sigma$-model for fluctuations of $u$ (derived in Ref.~\onlinecite{emp}) as well as
a more recent $\mathrm{O}(6)$-model \cite{sachdev13b} show in many aspects a good agreement with experiments.

\subsection{Antinodal states} 

The nontrivial order parameter~(\ref{k0}) has been derived taking into account only interactions mediated by
the critical modes with momenta $\sim\mathbf{Q}$ corresponding to the strongest antiferromagnetic fluctuations. Fermi surface regions beyond the hotspots 
have not yet been touched by the theoretical treatment and remained gapless in the ``hotspot-only'' approximation.

For the superconducting suborder, however, it is clear that the gap should cover the entire Fermi surface \cite{norman}, with the exception of
the nodes of the $d$-wave gap function that are situated at the intercept points of the Fermi surface and the diagonals of the Brillouin zone. 
In particular, we expect a significant superconducting gap also at the so-called \emph{antinodes} situated at the
zone edges, see Fig.~\ref{fig01}(b). The main result of the present study is that superconductivity is not the only possible order close to the antinodes, 
and we are going to show that another charge order (CDW) can appear in this region and challenge superconductivity there. Favorably for CDW,
opposite antinodes are effectively nested for a singular interaction. Flatness of the antinodal Fermi surface is not requisite but may enhance this effect.

To be specific, we extend the study of the spin-fermion model by considering both hotspots and antinodes, see Fig.~\ref{fig01}(b).
In the leading approximation, fermion quasiparticles located close to the antinodes interact with
hotspot fermions by exchanging non-singular paramagnons with propagator
$\langle\phi^{j}\phi^{j}\rangle \simeq [(\Delta K)^{2} + \xi_{\mathrm{AF}}^{-2}]^{-1}$ where $\Delta K = |\mathbf{K}_1-\mathbf{Q}|$
is the distance between hotspots and nearest antinodes, cf. Fig.~\ref{fig01}.
This interaction is clearly weaker than the interaction between hotspots connected by~$\mathbf{Q}$.
On the other hand, it allows quantum criticality to spread into the so far untouched antinodal
regions. The smallness of the non-singular propagators
justifies a perturbative treatment, whereas singular paramagnons have to be fully accounted for.

We now discuss the effects due to the paramagnon-mediated interaction between hotspot and antinodal quasiparticles.
We begin with the case of established superconductivity at the hotspots
and then, more interestingly, for the case of hotspots gapped by QDW or pseudogapped hotspots.

Below $T_c$, hotspot fermions form
Cooper pairs. In this case, we may neglect fluctuations and replace pairs of hotspot
fermion fields by their mean-field average $\Delta_\mathrm{SC} \sim b_0\langle\psi^\dagger_{\uparrow,1}\psi^\dagger_{\downarrow,3}\rangle$.
Let us consider two antinodal quasiparticles situated, e.g., at antinodes~$\mathbf{A}$ and $\mathbf{B}$, see Fig.~\ref{fig01}(c). Virtually
exchanging a paramagnon with wave vector~$\mathbf{K}_1$, they are scattered to hotspots~$\mathbf{1}$ and~$\mathbf{3}$. There,
they are affected by the established superconducting order~$\Delta_\mathrm{SC}$ and thus form Cooper pairs themselves. Interestingly, a similar virtual
process is impossible for the particle-hole suborder (QDW) since in this case both particle and hole
would have to emit a paramagnon of the same wave vector.

As a result, the explicit mean-field analysis, cf. Eq.~(\ref{026}), yields a superconducting
gap at the antinodes, which by a factor of
\begin{align}
\alpha \sim \frac{\Gamma^2}{v^2\big[(\Delta K)^2 + \xi_\mathrm{AF}^{-2}]} \label{k01}
\end{align}
is smaller than the hotspot gap. Notably, cf. again Eq.~(\ref{026}), the antinodal gap has $d$-wave symmetry. Moreover,
by continuity the antinodal superconductivity fixes the relative phase of the so far decoupled
superconducting suborders of the two hotspot quartets in Fig.~\ref{fig01}(a), ensuring
overall $d$-wave symmetry of the superconducting order parameter.
We note that this mean-field result actually does not require separating the Fermi surface
into hotspots and antinodal regions and has been obtained with full momentum resolution. \cite{norman}

\subsection{Charge density wave}

When hotspot superconductivity is destroyed by either thermal fluctuations
or a strong magnetic field, the superconducting gap at the hotspots has zero mean, $\langle\Delta_{\mathrm{SC}}\rangle=0$, 
implying absence of antinodal superconductivity as well. However, antinodal quasiparticles still couple
to non-zero superconducting fluctuations~$\Delta_{\mathrm{SC}}(\mathbf{r},\tau)$ induced at the antinodes
by the same mechanism that produced the antinodal superconducting gap in the preceding section.

In this situation, the superconducting fluctuations mediate an effective
interaction between antinodal fermions. Close to the transition,
the mass~$\xi^{-2}_{\mathrm{SC}}$ of superconducting fluctuations is small and 
the effective interaction becomes critical. This also leads to effective nesting of opposite antinodes.
As a result, this situation is remarkably
similar to the initial situation of hotspot fermions interacting via critical paramagnons.
While quantum-critical paramagnons reorganize the ground state of hotspot quasiparticles into the pseudogap state, 
the critical superconducting fluctuations 
play a very similar role at the antinodes and
trigger in analogy a transition to another phase. This repeated
triggering of orders thus constitutes a \emph{cascade} of phase transitions.

The order parameter formed at the antinodes is pure particle-hole pairing.
It cannot be a form of superconductivity because it has to be ``orthogonal'' to
the superconducting fluctuations that mediate the effective interaction.
Furthermore, particle-hole pairing at antinodes~$\mathbf{A}$ and~$\mathbf{B}$, see Fig.~\ref{fig01}(b),
is independent from particle-hole pairing at~$\mathbf{C}$ and~$\mathbf{D}$. This can be seen
as, e.g., wave vectors $\mathbf{K}_1$ and $\mathbf{K}_2$
mediate interactions at antinodes~$\mathbf{A}$ and~$\mathbf{B}$ but have no meaning for $\mathbf{C}$ and~$\mathbf{D}$,
where involved paramagnons carry wave vectors~$\mathbf{K}_3$ and $\mathbf{K}_4$.
Invariance under rotations of~$90^\circ$ then inevitably leads
to a bidirectional \emph{charge density wave} (CDW) order at the antinodes.

\begin{figure}[t]
\centerline{\includegraphics[width=\linewidth]{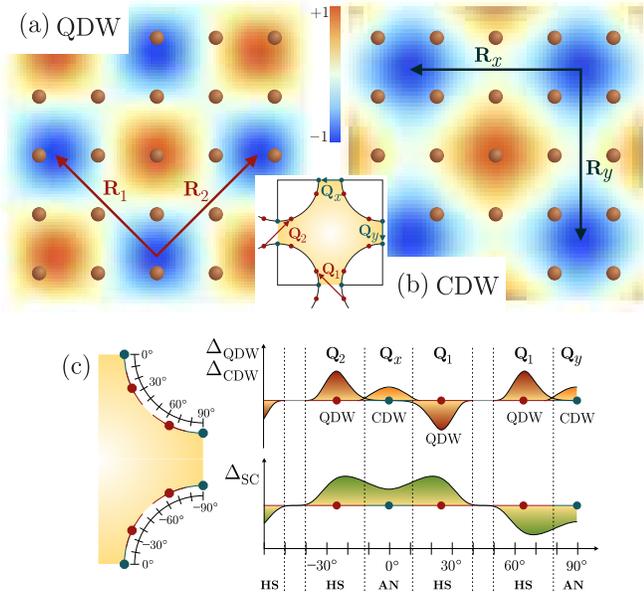}}
\caption{Checkerboard charge order for the pseudogap suborder
of (a) QDW and (b) antinodal CDW. Modulation vectors~$\mathbf{Q}_i$ giving
the periods $\mathbf{R}_{i}=2\protect\pi\mathbf{Q}_{i}/|\mathbf{Q}_{i}|^2$ are shown
in the inset. (c) Qualitative dependence of the superconducting and
charge order gaps on the position on the Fermi surface (HS = hotspots, AN = antinodes).}
\label{fig02}
\end{figure}

The explicit analysis (see Sec.~\ref{sec:micro}) follows
the same steps as the mean-field scheme of Ref.~\onlinecite{emp} for the pseudogap state.
This leads us to a similar universal mean-field equation, see Eq.~(\ref{045}), with all relevant energies measured
in units of the energy
\begin{align}
\Gamma_{\mathrm{CDW}}\sim\alpha^2\Gamma
\label{h01}
\end{align}
with~$\alpha \ll 1$ defined in Eq.~(\ref{k01}). A non-zero CDW gap
exists at temperatures $T<T_{\mathrm{CDW}} \sim 0.1\Gamma _{\mathrm{CDW}}$. 
In realistic cuprate systems, we may expect $T_{c}<T_{\mathrm{CDW}}<T^*$ as well as 
comparable energy scales, $\Gamma_{\mathrm{CDW}}\sim\Gamma$.
The calculation of charge density~$\rho(\mathbf{r})$ in the CDW phase leads to a spatial modulation
of the form
\begin{align}
\rho_{\mathrm{CDW}}(\mathbf{r})
\sim \frac{e\Gamma^2_{\mathrm{CDW}}}{v^2}
\big\{
   \cos(\mathbf{Q}_{x}\mathbf{r}+\varphi_{x})
  +\cos(\mathbf{Q}_{y}\mathbf{r}+\varphi_{y})
\big\}\ .
\label{c02}
\end{align}
The wave vectors~$\mathbf{Q}_{x}$ and $\mathbf{Q}_{y}$ (see Fig.~\ref{fig02})
connect opposite antinodes and correspond to a \emph{modulation
along the bonds} of the Cu lattice. The resulting pattern is a checkerboard
as shown in Fig.~\ref{fig02}(b), similarly to the pattern of QDW shown in Fig.~\ref{fig02}(a). 
Notably, the CDW and QDW patterns are
turned by $45^{\circ}$ with respect to each other.
Variables~$\varphi_{x,y}$ denote offset phases.
Figure~\ref{fig02}(c) summerizes the results of our study by providing a sketch of the emergent orders as a 
function of the position on the Fermi surface.

\section{Microscopic analysis}
\label{sec:micro}

\subsection{Effective Lagrangian}

We begin our microscopic analysis by developing a convenient and compact notation for the subsequent calculations.
We are mainly interested in the low-lying excitations close to the hotspots and antinodes, which
we numerate according to Fig.~\ref{fig01}(b) with numbers~$j=1,\ldots,8$ and capital letters~$J=\mathrm{A},\ldots,\mathrm{D}$, respectively. 
In this spirit, we represent a general quasiparticle field~$\psi(\mathbf{r})$ as
\begin{align}
\psi(\mathbf{r}) &= \sum_{j=1}^8 \mathrm{e}^{\mathrm{i}\mathbf{p}_j\mathbf{r}}\psi_j(\mathbf{r})
+ \sum_{J=\mathrm{A}}^{\mathrm{D}} \mathrm{e}^{\mathrm{i}\mathbf{p}_J\mathbf{r}}\chi_J(\mathbf{r})
\label{h11}\ ,
\end{align}
where~$\mathbf{p}_j$ and~$\mathbf{p}_J$ denote the positions of hotspots~$j$ and antinodes~$J$, respectively,
in the Brillouin zone. The fields for hotspot quasiparticles~$\psi_j$ and for antinodal ones~$\chi_J$
fluctuate only slowly in space on scales much larger than the lattice constant~$a$.

Following Ref.~\onlinecite{emp}, we introduce three pseudospin sectors $\mathrm{L}\otimes\Lambda\otimes\Sigma$ to organize the hotspot
states,
\begin{align}
\boldsymbol{\psi} =
\left(
\begin{array}{c}
  \left(
    \begin{array}{c}
       \left(
         \begin{array}{c}
           \psi_{1}\\
           \psi_{2}
         \end{array}
       \right)_\Sigma
       \\
       \left(
         \begin{array}{c}
           \psi_{3}\\
           \psi_{4}
         \end{array}
       \right)_\Sigma
    \end{array}
  \right)_\Lambda\\
  \left(
    \begin{array}{c}
       \left(
         \begin{array}{c}
           \psi_{5}\\
           \psi_{6}
         \end{array}
       \right)_\Sigma
       \\
       \left(
         \begin{array}{c}
           \psi_{7}\\
           \psi_{8}
         \end{array}
       \right)_\Sigma
    \end{array}
  \right)_\Lambda
\end{array}
\right)_\L
\label{002}\ .
\end{align}
Inspecting the structure defined in Eq.~(\ref{002}), we see that the sector~$\L$
organizes the hotspots in the two quartets along the diagonals of the Brillouin zone,
cf. Fig.~\ref{fig01}(a). Sector~$\Lambda$ distinguishes inside each of the
quartets the two pairs of hotspots connected by the antiferromagnetic ordering wave
vector~$\mathbf{Q}$. Finally, the pseudospin~$\Sigma$ corresponds to the
two hotspots within each of such pairs. The antinodal fields are similarly 
combined into
\begin{align}
\boldsymbol{\chi} =
\left(
\begin{array}{c}
  \left(
    \begin{array}{c}
       \chi_{\mathrm{A}}\\
       \chi_{\mathrm{B}}
    \end{array}
  \right)_\Upsilon\\
  \left(
    \begin{array}{c}
       \chi_{\mathrm{C}}\\
       \chi_{\mathrm{D}}
    \end{array}
  \right)_\Upsilon
\end{array}
\right)_\Xi
\label{001}\ ,
\end{align}
where~$\Xi$ and~$\Upsilon$ are two more pseudospins for the four antinodes
in Fig.~\ref{fig01}(b). Operators acting on these various pseudospin spaces are conveniently expanded in
Pauli matrices denoted by, e.g., $\Upsilon_1$ for the first Pauli matrix
in $\Upsilon$ space. Each of the field components~$\psi_j$ and~$\chi_J$ in Eqs.~(\ref{002}) and~(\ref{001})
is itself a spinor for the physical spin, for which we use as usual the Pauli matrix notation $\boldsymbol{\sigma}=(\sigma_1,\sigma_2,\sigma_3)$.

In the approximation of linearized Fermi surfaces close to hotspots and antinodes, 
the non-interacting part~$\mathcal{L}_0$ of the Lagrangian reads
\begin{align}
\mathcal{L}_0 =
\boldsymbol{\chi}^\dagger
    \big(
       \partial_\tau
       - \mathrm{i}\hat{\mathbf{v}}\nabla
    \big)
\boldsymbol{\chi}
+
\boldsymbol{\psi}^\dagger
    \big(
       \partial_\tau
       - \mathrm{i}\hat{\mathbf{V}}\nabla
    \big)
\boldsymbol{\psi}
\ ,\label{i11}
\end{align}
where the velocity operator for the antinodal states reads
\begin{align}
\hat{\mathbf{v}} = - \frac{v}{2}\big[
\Upsilon_3(1+\Xi_3)\mathbf{e}_x
+
\Upsilon_3(1-\Xi_3)\mathbf{e}_y
\big]
\ .\label{002d}
\end{align}
Herein, $\mathbf{e}_x$ and $\mathbf{e}_y$ are unit vectors in the directions
of Cu bonds and $v$ is the (antinodal) Fermi velocity. The hotspot velocity operator~$\hat{\mathbf{V}}$ is a little more complicated. Since we do not use this operator in the present study directly, we refer the reader to Ref.~\onlinecite{emp}.

During the analysis, it will be convenient to study charge and superconducting
correlations on equal footing. Therefore, we introduce another pseudospin~$\tau$
distinguishing particle and hole states,
\begin{align}
\Psi = \frac{1}{\sqrt{2}}
\left(
\begin{array}{c}
\boldsymbol{\psi}\\
\mathrm{i}\sigma_2\boldsymbol{\psi}^*
\end{array}
\right)_\tau
\ , \quad
\X = \frac{1}{\sqrt{2}}
\left(
\begin{array}{c}
\boldsymbol{\chi}\\
\mathrm{i}\sigma_2\boldsymbol{\chi}^*
\end{array}
\right)_\tau
\label{002g}\ .
\end{align}
The matrix $C = -\tau_2\sigma_2$
allows for a definition of charge-conjugation
\begin{align}
\bar{\Psi} = \Psi^\mathrm{t} C 
\ , \quad
\bar{\X} = \X^\mathrm{t} C 
\label{002i}\ .
\end{align}
In particle-hole space notation, the Lagrangian~(\ref{i11}) becomes
\begin{align}
\mathcal{L}_0 &=
-\bar{\X}
    \big(
       \partial_\tau
       - \mathrm{i}\hat{\mathbf{v}}\nabla
    \big)
\X
-\bar{\Psi}
    \big(
       \partial_\tau
       - \mathrm{i}\hat{\mathbf{V}}\nabla
    \big)
\Psi
\label{i21}\ ,
\end{align}
which concludes the non-interacting part of the effective theory.\bigskip

In order to incorporate the interaction mediated by paramagnons~$\boldsymbol{\phi}$ into the model, 
we again single out the relevant modes. These are those harmonics of the field~$\boldsymbol{\phi}$
with wave vector close to~$\mathbf{Q}$ for hotspot--hotspot interaction and wave vectors at~$\mathbf{K}_1,\ldots,\mathbf{K}_4$
for hotspot--antinode interactions, see Fig.~\ref{fig01}(b). We assume that~$\mathbf{K}_1-\mathbf{K}_2$
is not an inverse lattice vector, which for a general curved Fermi surface is the correct assumption. 

In the compact notation, the Lagrangian for interaction at wave vectors~$\sim\mathbf{Q}$ is written as\cite{emp}
\begin{align}
\mathcal{L}_{\mathrm{int},\mathbf{Q}} &= \lambda\ \bar{\Psi}\Sigma_1(\boldsymbol{\phi}_0\boldsymbol{\sigma})\Psi\ .
\label{hm012}
\end{align}
Here $\lambda$ is the coupling constant for the paramagnon--fermion interaction. 
The general correlation function for $\boldsymbol{\phi}$, Eq.~(\ref{a01}), translates
to the correlation
\begin{align}
\big\langle\phi^\alpha_{0,\omega,\mathbf{q}}\phi^\beta_{0,-\omega,-\mathbf{q}} \big\rangle
 &= 
\frac{\delta_{\alpha\beta}}{c^{-2}\omega^2+\mathbf{q}^2+\xi_{\mathrm{AF}}^{-2}}
\label{hm013}
\end{align}
for the field~$\boldsymbol{\phi}_0$ entering Eq.~(\ref{hm012}). 

For the interaction at wave vectors 
$\mathbf{K}_1,\ldots,\mathbf{K}_4$, we introduce fields~$\boldsymbol{\phi}_{\pm k}$ that are
related to field~$\boldsymbol{\phi}$ of Eq.~(\ref{a01}) as
\begin{align}
\boldsymbol{\phi}_{\pm k,\mathbf{q},\omega} &= \boldsymbol{\phi}_{\pm\mathbf{K}_k+\mathbf{q},\omega}
\label{hm014}
\end{align}
with correlations
\begin{align}
\big\langle\phi^\alpha_{k,\omega,\mathbf{q}}\phi^\beta_{-k,-\omega,-\mathbf{q}} \big\rangle
 &\simeq 
\frac{\delta_{\alpha\beta}}{(\Delta K)^2+\xi_{\mathrm{AF}}^{-2}}
\ .  \label{h013}
\end{align}
While hotspot--hotspot paramagnons~$\boldsymbol{\phi}_{0}$ become critical at the antiferromagnetic QCP ($\xi_{\mathrm{AF}}\rightarrow\infty$),
paramagnons~$\boldsymbol{\phi}_{\pm k}$ are effectively static as $\Delta K=|\mathbf{K}_1-\mathbf{Q}|\gg |\mathbf{q}|,c^{-1}\omega$
at low energies. The corresponding Lagrangian reads
\begin{align}
\mathcal{L}_{\mathrm{int},\mathbf{K}} &=
2\lambda \sum_{k=1}^4\Big( 
\bar{\Psi} T_k(\boldsymbol{\phi}_k\boldsymbol{\sigma}) \X
+
\bar{\X}\ T_k^\mathrm{t}(\boldsymbol{\phi}_k\boldsymbol{\sigma})\Psi
\Big)
\ ,
\label{006}
\end{align}
where matrices~$T_k$ describe the various scattering processes between hotspots and antinodes.
They are given by
\begin{align}
T_1 &=
\left(
\begin{array}{cc}
t_{3}^{\mathrm{B}} & 0 \\
0 & t_{1}^{\mathrm{A}}
\end{array}
\right)_\tau\ ,\quad
T_2 =
\left(
\begin{array}{cc}
t_{6}^{\mathrm{A}} & 0 \\
0 & t_{8}^{\mathrm{B}}
\end{array}
\right)_\tau\ ,\nonumber\\
T_3 &=
\left(
\begin{array}{cc}
t_{7}^{\mathrm{D}} & 0 \\
0 & t_{5}^{\mathrm{C}}
\end{array}
\right)_\tau\ ,\quad
T_4 =
\left(
\begin{array}{cc}
t_{4}^{\mathrm{C}} & 0 \\
0 & t_{2}^{\mathrm{D}}
\end{array}
\right)_\tau \ ,
\label{007}
\end{align}
where the $8\times 4$ matrices~$t_j^J$ are defined by $\boldsymbol{\psi}^\dagger t_j^J \boldsymbol{\chi}=\psi_j^\dagger\chi_J$.
While non-trivial effects due to the hotspot Lagrangian~$\mathcal{L}_{\mathrm{int},\mathbf{Q}}$, Eq.~(\ref{hm012}), have been extensively studied
in Refs.~\onlinecite{ms2,emp}, we are now in a position to extend the physical picture by effects emerging in the antinodal region, which
couple nontrivially to the hotspots via Lagrangian~$\mathcal{L}_{\mathrm{int},\mathbf{K}}$, Eq.~(\ref{006}).

\subsection{Emerging orders}

\subsubsection{Pseudogap state}

Coupling between hotspot fermions and quantum-critical paramagnons~$\boldsymbol{\phi}_0$, Eq.~(\ref{hm013}),
has been studied for a long time. In Ref.~\onlinecite{emp}, it was shown that close to the QCP ($\xi_{\mathrm{AF}}\rightarrow\infty$) below
a temperature~$T^*\sim\Gamma\sim\lambda^2$, an unusual order
parameter composed of two competing suborders appears. These are superconductivity with
complex amplitudes $\Delta_\mathrm{SC}^1$ and $\Delta_\mathrm{SC}^2$ and a charge order
of a spatially modulated quadrupole moment (quadrupole density wave, QDW) with amplitudes
$\Delta_\mathrm{QDW}^1$ and $\Delta_\mathrm{QDW}^2$, cf. Eq.~(\ref{k0}). Upper indices refer to the
two decoupled quartets of hotspots given by $\L=1$ and $\L=2$ states, respectively, cf. Fig.~1(a).
This order, hereafter referred to as ``pseudogap'', constitutes a stable saddle-point manifold in
the theory~$\mathcal{L}_0+\mathcal{L}_{\mathrm{int},\mathbf{Q}}$. We incorporate it
in terms of a mean-field term that replaces~$\mathcal{L}_{\mathrm{int},\mathbf{Q}}$, Eq.~(\ref{hm012}),
in the model. This term is given by
\begin{align}
\mathcal{L}_\mathrm{PG} &=
\bar{\Psi}\ 
b(\mathrm{i}\partial_\tau) \mathcal{O}_\mathrm{PG}
\Psi
\ ,\label{i22}
\end{align}
where $b(\varepsilon)$ is a function of fermionic Matsubara frequencies~$\varepsilon$ and $\mathcal{O}_\mathrm{PG}$
is a matrix in the pseudospin spaces that reflects the symmetry of the order parameter.
It reads\cite{emp}
\begin{align}
\mathcal{O}_\mathrm{PG} &=\mathrm{i}\Sigma_3
\left(
\begin{array}{cc}
\left(
\begin{array}{cc}
0 & u_1\\
- u_1^\dagger & 0
\end{array}
\right)_\Lambda & 0
\\
0 &
\left(
\begin{array}{cc}
0 & u_2\\
- u_2^\dagger & 0
\end{array}
\right)_\Lambda
\end{array}
\right)_{\L}\ .
\label{024}
\end{align}
Here, $u_1$ and $u_2$ are $\mathrm{SU}(2)$ matrices in particle-hole space for each of the two quartets of hotspots.

Let us expand the~$u_j$ in particle-hole space Pauli matrices~$\tau_i$,
\begin{align}
u_j = \Delta^j_0 + \mathrm{i}\big(
\Delta^j_1 \tau_1+\Delta^j_2 \tau_2+\Delta^j_3 \tau_3
\big)
\ ,\label{025}
\end{align}
so that $\Delta^j_{\mathrm{QDW}}=\Delta^j_0+\mathrm{i}\Delta^j_3$ and $\Delta_{\mathrm{SC}}^j=\Delta^j_1+\mathrm{i}\Delta^j_2$.
Numbers $\Delta^j_n$ are real and satisfy the constraint $\sum_{n=0}^3 [\Delta^j_n]^2=1$ imposed by unitarity. At low energies,
we may approximate \cite{emp} the function~$b(\varepsilon)$ as a (positive) constant, $b(\varepsilon)\simeq b_0$.

Study of fluctuations \cite{emp} of the pseudogap $b(\varepsilon)\mathcal{O}$ shows that below a temperature $T_c<T^*$,
one of the suborders ---QDW or superconductivity--- is suppressed, 
provided symmetry-breaking effects such as curvature of the Fermi surface are included in the consideration. 
In the absence of the magnetic field, finite
curvature makes the composite order parameter prefer superconductivity as the ground state, whereas a sufficiently strong magnetic field can make a charge modulated state (QDW) energetically more favourable.\cite{mepe} Between $T_c$ and $T^*$, neither are capable of forming a long-range order and the system is in a regime of strong thermal fluctuations
between the two suborders.

\subsubsection{Antinodal superconductivity}

Averaging the Lagrangian~(\ref{006}) over the paramagnon fluctuations~$\boldsymbol{\phi}_k$, Eq.~(\ref{h013}), 
yields an effective $4$-point interaction vertex
\begin{align}
\mathcal{L}_\mathrm{int}
&=
-\frac{4\lambda^2}{(\Delta K)^2}\sum_{k=1}^4
\bar{\X}\tau_1 T_k^\mathrm{t}\tau_1\boldsymbol{\sigma}\Psi\bar{\Psi}\boldsymbol{\sigma}\tau_1T_k\tau_1\X \ .
\label{022}
\end{align}
The model $\mathcal{L}_0+\mathcal{L}_\mathrm{PG}+\mathcal{L}_\mathrm{int}$, Eqs.~(\ref{i21}), (\ref{i22}), and (\ref{022}),
is the effective model our subsequent study on the physics at the antinodes is based on.

In a mean-field scheme to decouple the interaction $\mathcal{L}_\mathrm{int}$, Eq.~(\ref{022}),
we replace the $\Psi\bar{\Psi}$ operator by its mean-field correlation function,
which by Eqs~(\ref{i21}) and~(\ref{i22}) is given by
\begin{align}
\langle\Psi\bar{\Psi}\rangle_{\mathrm{m.f.}} = \frac{J(T)}{4\pi}\ \mathcal{O}\ .
\label{023}
\end{align}
The function~$J(T)$ is defined as
\begin{align}
J(T)= \frac{\Omega T}{v}\sum_{\varepsilon}
\frac{b(\varepsilon)}
{\sqrt{\varepsilon^2+b^2(\varepsilon)}}
\label{023a}
\end{align}
and $\Omega \sim \lambda^2/v$ is the volume of the hotspot, cf. Ref.~\onlinecite{emp}.
Inside the pseudogap regime, the function~$J(T)\sim \lambda^4/v^2$
is in a good approximation independent of the temperature~$T$, while it turns to zero when $T$ approaches~$T^*$.

Inserting Eq.~(\ref{023}) into  Eq.~(\ref{022}) yields $\mathcal{L}_\mathrm{int}\simeq \mathcal{L}_{\mathrm{m.f.}}$
with the mean-field Lagrangian given by
\begin{align}
\mathcal{L}_{\mathrm{m.f.}}&=
 \frac{3\lambda^2J(T)}{\pi (\Delta K)^2}
\big[
\bar{\X}\
\Xi_3\Upsilon_1
\big\{
\Delta_1
\tau_1 +
\Delta_2
\tau_2
\big\}\ \X
\big]\ .
\label{026}
\end{align}
Herein, $\Delta_1=(\Delta_1^1-\Delta_1^2)/2$ and $\Delta_2=(\Delta_2^1-\Delta_2^2)/2$ form the effective
amplitude $\Delta_{\mathrm{SC}}=\Delta_1 + \mathrm{i}\Delta_2$ of the hotspot superconductivity.
Importantly, in this mean-field treatment, only the superconducting suborder of
the hotspot pseudogap gives a contribution, while the QDW does not effectively couple
to the fields $\bar{\X}$ and~$\X$ so that it does not play a direct role at the antinodes.
Equation~(\ref{026}) thus demonstrates that the hotspot superconductivity induces
a superconducting order parameter at the antinodes by the same mechanism sketched in Fig.~\ref{fig01}(c) and discussed in
Sec.~\ref{sec:picture}. The presence of~$\Xi_3$ reflects
the $d$-wave symmetry of the superconducting order. The order parameter of antinodal superconductivity is
maximal if
\begin{align}
\Delta_{1,2}^1 = -\Delta_{1,2}^2\ ,
\label{027}
\end{align}
which should be energetically the favoured configuration. Note that the matching condition~(\ref{027})
reduces the $\mathrm{O}(4)\times\mathrm{O}(4)$ symmetry of the hotspot order to a constrained $\mathrm{O}(6)$ model,
cf. Ref.~\onlinecite{sachdev13b}.

Let us estimate the strength of the superconducting gap induced at the antinodes.
According to Eq.~(\ref{023a}), we estimate $J(T)$ inside the pseudogap
as~$J(T)\sim \lambda^4/v^2$, which is smaller
than the high energy scale given by the momentum distance~$\Delta K$ between hotspots and antinodes.
Thus, while the hotspot pseudogap is of order~$\Gamma\sim\lambda^2$, the induced antinodal superconducting gap
is of order $\lambda^2 [\lambda^4/(v\Delta K)^2]\sim \alpha \Gamma \ll \Gamma$, cf. Eq.~(\ref{k01}). We emphasize once more that the
antinodal superconductivity is induced only if the hotspot system is in the
superconducting state.

\subsubsection{Antinodal charge-density wave order}

Let us now address the case when hotspot superconductivity is destroyed by either thermal fluctuations above~$T_c$ (pseudogap state) 
or by a strong enough magnetic field at arbitrary temperature. 
In the latter case, we obtain QDW at $T<T_c$ or the pseudogap state at $T>T_c$ 
instead of the superconductor.
Then, the mean-field decoupling in Eq.~(\ref{026}) does not induce a
finite gap at the antinodes as~$\Delta_1=\Delta_2=0$. However, superconducting fluctuations are still present
even if $\langle\Delta_1(\mathbf{r},\tau)\rangle=\langle\Delta_2(\mathbf{r},\tau)\rangle=0$.
These fluctuations have been studied with the help of a non-linear $\sigma$-model in Ref.~\onlinecite{emp}.

At not too high temperatures above the superconducting critical temperature $T_c$ at zero field or below $T_c$ in a sufficiently strong 
magnetic field destroying the superconductivity, the superconducting fluctuations $\Delta_\mathrm{SC}(\mathbf{r},\tau)$ are small and the $\sigma$-model yields 
the effective Lagrangian
\begin{align}
\mathcal{L}_{\mathrm{fluct}} \simeq \dfrac{g\lambda^2}{2}
\big(
|\partial_\mu\Delta_\mathrm{SC}|^2 + \xi^{-2}_{\mathrm{SC}}\ |\Delta_\mathrm{SC}|^2
\big)
\label{041}
\end{align}
with $\partial_\mu = (u^{-1}\partial_\tau,\nabla)$, $g\sim 1$ a coupling constant, and $u\sim v$ the velocity of the fluctuation modes.
For $T>T_c$ it is not easy to carry out explicit calculations in the pseudogap state. However, it is well-known\cite{zinn} that there is no phase transition in the two-dimensional fully isotropic O(4)-symmetric $\sigma$-model as all excitations have a gap. In our situation this means that correlation functions of superconducting fluctuations can still formally be obtained from Eq.~(\ref{041}) but the constants entering this equations have now to be considered as effective parameters whose values can hardly be calculated analytically.
In the subsequent analysis, we assume that the length $\xi_{\mathrm{SC}}$ diverges on the critical line separating the superconducting region from QDW or pseudogap phase.

In the Gaussian approximation of Eq.~(\ref{041}), we immediately integrate the fluctuation modes out of the Lagrangian~(\ref{026}) (, where
$\Delta_\mathrm{SC}=\Delta_1+\mathrm{i}\Delta_2$ is now assumed
to fluctuate both in space and time). Then, we obtain the effective interaction between the antinodal fermions,
\begin{align}
&\mathcal{L}_{\mathrm{int},\mathrm{fluct}}
=
 -\frac{9\lambda^2 J^2(T)}{\pi^2 g (\Delta K)^4}
\sum_{j=1}^{2}
\big(
\bar{\X}(\mathbf{r},\tau)\
\Xi_3\Upsilon_1
\tau_j
\ \X(\mathbf{r},\tau)
\big)\nonumber\\
&\times \Phi(\mathbf{r}-\mathbf{r}',\tau-\tau')
\big(
\bar{\X}(\mathbf{r}',\tau')\
\Xi_3\Upsilon_1
\tau_j
\ \X(\mathbf{r}',\tau')
\big)
\ ,
\label{043}
\end{align}
where
\begin{align}
 \Phi_{\mathbf{q},\omega}  =
 \frac{1}{u^{-2}\omega^2 + \mathbf{q}^2+\xi^{-2}_{\mathrm{SC}}}
\label{044}
\end{align}
is the propagator of superconducting fluctuations. At the transition, $\xi_\mathrm{SC}\rightarrow\infty$ and this propagator is singular in the infrared limit, which makes
the antinodal points effectively hot. Moreover, opposite antinodes are effectively nested. We emphasize, though, that, in analogy
with the hotspot fermions interacting via critical paramagnons, this effective nesting is due to the singular form of the propagator of superconducting fluctuations in the vicinity of the superconductor transition where the length $\xi_{\mathrm{SC}}$ diverges. This does not necessarily require a geometrically flat Fermi surface at the antinodes.

\emph{The interaction~(\ref{043}) generates an instability toward charge-density wave (CDW) order.}
Indeed, the Lagrangian~(\ref{043}) for the interaction of antinodal fermions
has effectively the same form as the effective interaction induced by paramagnons that is responsible
for the formation of the pseudogap. 
We thus introduce a CDW order parameter in the Lagrangian,
\begin{align}
\mathcal{L}_\mathrm{CDW} &= \bar{\X}b_{\mathrm{CDW}}(\mathrm{i}\partial_\tau)\mathcal{O}_\mathrm{CDW}\X
\label{hm031}\ ,
\end{align}
and obtain in analogy with Ref.~\onlinecite{emp} the mean-field equation
\begin{widetext}
\begin{align}
b_{\mathrm{CDW}}(\varepsilon)\mathcal{O}_\mathrm{CDW}
=
 -\frac{9\lambda^2 J^2(T)}{\pi^2 g (\Delta K)^4}
\sum_{j=1}^{2}T\sum_{\varepsilon',\mathbf{k}'}
 \Phi_{\mathbf{k}',\varepsilon-\varepsilon'}
\Xi_3\Upsilon_1
\tau_j
\ 
\frac{b_{\mathrm{CDW}}(\varepsilon')\mathcal{O}_\mathrm{CDW}}{\varepsilon'^2+(v\mathbf{k}')^2+b^2_{\mathrm{CDW}}(\varepsilon')}
\
\Xi_3\Upsilon_1
\tau_j
\ .
\label{hm032}
\end{align}
\end{widetext}
Deriving Eq.~(\ref{hm032}) has required that~$\mathcal{O}_\mathrm{CDW}$ anticommutes
with the velocity operator~$\hat{\mathbf{v}}$, Eq.~(\ref{002d}), which implies
$\{\mathcal{O}_\mathrm{CDW},\Upsilon_3\}=0$ and $[\mathcal{O}_\mathrm{CDW},\Xi_3]=0$.
In addition, we assume the normalization~$\mathcal{O}_\mathrm{CDW}^2=1$.
Furthermore, in order to compensate for the minus sign in Eq.~(\ref{hm032}), 
we need to impose that~$\{\mathcal{O}_\mathrm{CDW},\Upsilon_1\tau_1\}=0$ and 
$\{\mathcal{O}_\mathrm{CDW},\Upsilon_1\tau_2\}=0$. Summarizing all these constraints,
the antinodal order parameter becomes
\begin{align}
\mathcal{O}_\mathrm{CDW} =
\left(
\begin{array}{cc}
\Delta_x' \Upsilon_1\tau_3 + \Delta_x'' \Upsilon_2 & 0 \\
0 & \Delta_y' \Upsilon_1\tau_3 + \Delta_y'' \Upsilon_2
\end{array}
\right)_\Xi
\label{044}\ .
\end{align}
Parameters $\Delta_x'$ and $\Delta_x''$ play the roles of real and imaginary parts for the order parameter of CDW in $x$-direction
while $\Delta_y'$ and $\Delta_y''$ do so for the $y$-direction. They satisfy the nonlinear
constraints $[\Delta_x']^2+[\Delta_x'']^2=1$ and $[\Delta_y']^2+[\Delta_y'']^2=1$.

\begin{figure}[t]
\centerline{\includegraphics[width=0.7\linewidth]{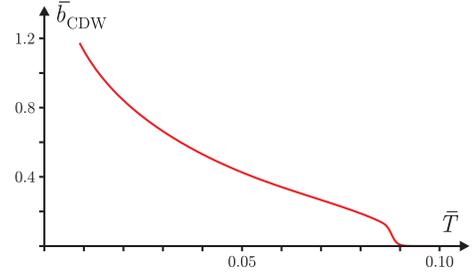}}
\caption{Dimensionless charge-density gap $\bar{b}_\mathrm{CDW}=b_\mathrm{CDW}/\Gamma_\mathrm{CDW}$ as a function of dimensionless temperature~$\bar{T}$ interpolated to the frequency $\varepsilon=0$. A CDW order appears below the temperature~$T_\mathrm{CDW}\approx 0.09\Gamma_\mathrm{CDW}$.
}
\label{fig06}
\end{figure}

Measuring all quantities of dimension of energy in units of
\begin{align}
\Gamma_{\mathrm{CDW}}&=\frac{18u\lambda^2 J^2}{\pi^2 g v(\Delta K)^4}
\label{hm041}\ ,
\end{align}
we derive from Eq.~(\ref{hm032}) a \emph{universal} self-consistency equation for
the CDW amplitude~$b_{\mathrm{CDW}}(\varepsilon)$,
\begin{align}
\bar{b}_{\mathrm{CDW}}(\varepsilon)
&= \bar{T}\sum_{\bar{\varepsilon}'}
 \frac{1}{|\bar{\varepsilon}-\bar{\varepsilon}'|}
 \frac{\bar{b}_{\mathrm{CDW}}(\bar{\varepsilon}')}
           {\sqrt{\bar{\varepsilon}'^2 +  \bar{b}_{\mathrm{CDW}}^2(\bar{\varepsilon}')}}
           \ .
\label{045}
\end{align}
In this equation, all quantities~$z$ of dimension energy enter in the form~$\bar{z}=z/\Gamma_{\mathrm{CDW}}$,
The energy scale~$\Gamma_{\mathrm{CDW}}\sim \alpha^2 \Gamma$, cf. Eq.~(\ref{k01}), is smaller than both the pseudogap energy scale~$\sim\Gamma$ and
the antinodal superconducting gap $\sim \alpha\Gamma$, which appears when the pseudogap has ordered into
the superconducting suborder. Numerical investigation of Eq.~(\ref{045}) indicates non-zero solutions for~$b_{\mathrm{CDW}}(T,\varepsilon)$ below a temperature
$T_\mathrm{CDW} \approx 0.09 \Gamma_\mathrm{CDW}$. Figure~\ref{fig06} shows the (interpolated) amplitude~$b_{\mathrm{CDW}}(T,0)$ as a function
of temperature~$T$.

Calculating the charge density in the presence of the order parameter~$\mathcal{O}_\mathrm{CDW}$,
Eq.~(\ref{044}), we obtain formula~(\ref{c02}) for the bidirectional CDW modulation,
\begin{align}
\rho_{\mathrm{CDW}}(\mathbf{r})
\sim \frac{e\Gamma_{\mathrm{CDW}}^2}{v^2}
\big\{
\cos(\mathbf{Q}_x\mathbf{r}+\varphi_x)+
\cos(\mathbf{Q}_y\mathbf{r}+\varphi_y)
\big\}
\label{046}\ ,
\end{align}
where $\varphi_x$ and $\varphi_y$ denote the phases of the CDW order in $x$ and $y$ directions, respectively.
Thus, the charge density is modulated with the wave vectors $\mathbf{Q}_x$ and $\mathbf{Q}_y$ connecting
two opposite antinodal points. This contrasts the modulations of the quadrupole-density~$D_{xx}$ generated\cite{emp} at the hotspots
in the presence of QDW,
\begin{align}
D_{xx}(\mathbf{r}) &\sim e\big\{|\Delta_{\mathrm{QDW}}^1|\cos(\mathbf{Q}_1\mathbf{r}+\varphi_1)
\nonumber\\
&\qquad +
|\Delta_{\mathrm{QDW}}^2|\cos(\mathbf{Q}_2\mathbf{r}+\varphi_2)\big\}
\label{047}\ .
\end{align}
QDW wave vectors $\mathbf{Q}_1$ and $\mathbf{Q}_2$ are turned by $45^{\circ}$ and longer than the CDW
wave vectors by a factor roughly given by $\sqrt{2}$. Both orders form checkerboards as illustrated in
Fig.~\ref{fig02}. Figure~\ref{fig02}(c) shows the type of particle-hole order, i.e. whether QDW or CDW, as a qualitative function of the position
on the Fermi surface. Whereas within our model hotspot and antinodal regions are separated, we expect in realistic systems
regions of small overlap of the two orders in between.

\section{Cuprate physics}
\label{sec:cuprates}

\begin{figure}[t]
\centerline{\includegraphics[width=0.8\linewidth]{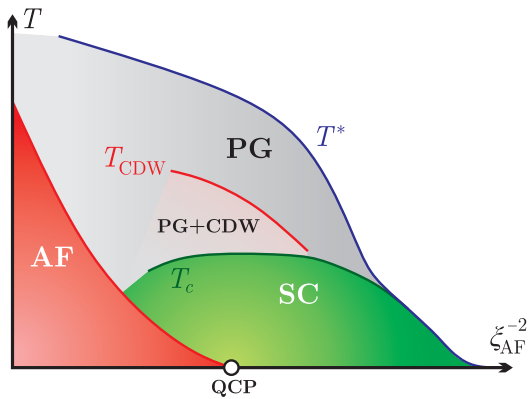}}
\caption{Qualitative phase diagram summarizing the results of Ref.~\onlinecite{emp} and the present work for zero magnetic field.
Close to the antiferromagnetic (AF) QCP, $\xi_{\mathrm{AF}}^{-2}=0$, and upon lowering the temperature, the systems develops first at~$T^*$ the instability toward
the fluctuating pseudogap state (PG) characterized by the order parameter of Eq.~(\ref{k0}). At lower temperatures~$T<T_\mathrm{CDW}<T^*$, strong superconducting fluctuations 
induce a transition toward charge density wave (CDW) formed at the antinodes. Finally, below~$T_c$, the particle-particle suborder 
of the pseudogap prevails due to curvature effects and establishes $d$-wave superconductivity.}
\label{fig07}
\end{figure}

We now address the phase diagram of cuprates in the proximity of
the antiferromagnetic QCP. We emphasize that our theory applies only to the
``metallic'' side of the antiferromagnet--normal metal phase transition. The regions of too low doping are thus excluded in the following
discussion. In the
region of intermediate doping, suppression of carrier density below a crossover temperature~$T^*$ observed in NMR measurements \cite{warren,alloul}
was the first evidence for the existence of a ``pseudogap'' in the electron spectrum. In contrast, $d$-wave superconductivity
appears only below a considerably lower temperature~$T_{c}$. In our theory, $T^*$ is associated with the crossover to the
strongly fluctuating $\mathrm{O}(4)$-symmetric composite order (superconductivity and QDW) close to the hotspots. \cite{emp}
The phase diagram, see Fig.~\ref{fig07}, is further enriched by the formation
of CDW order with wave vectors $\mathbf{Q}_{x,y}$ (Fig.~\ref{fig02}) at the edge of the
Brillouin zone. Also the emergence of the CDW order is ultimately due
to the proximity to the QCP. The additional phase transition is expected
to occur at a temperature~$T_{\mathrm{CDW}}$ inside the
pseudogap phase, $T_{c}<T_{\mathrm{CDW}}<T^*$.

The charge modulation observed in various recent experiments \cite{wise,davis,yazdani,julien,ghiringhelli,chang,achkar,leboeuf,blackburn}
has been attributed \cite{emp,sachdev13a} to the existence of
QDW (or ``bond order'') correlations. This picture is, in principle, in agreement
with NMR results \cite{julien} and sound propagation measurements \cite{leboeuf,shekhter}. However, STM studies \cite{wise,davis,yazdani}
of BSCCO and experiments with hard \cite{chang,blackburn} and resonant
soft \cite{ghiringhelli,achkar} X-ray scattering on YBCO have revealed a
charge modulation along the bonds of the Cu lattice with modulation
vectors close to $\mathbf{Q}_{x,y}$, which are the CDW wave vectors.
Moreover, QDW has a vanishing Fourier transform
near even Bragg peaks. Therefore, STM and hard X-ray experiments
can hardly be expected to detect the QDW modulation.

The seeming contradiction is resolved when we include the
CDW, Eq.~(\ref{c02}), in the Cu lattice. Then, this explains the experimental results \cite{wise,davis,yazdani,julien,ghiringhelli,chang,achkar,leboeuf,blackburn}.
CDW appears below a critical temperature~$T_{\mathrm{CDW}}$ that can be considerably lower than $T^*$,
in line with the results of the hard X-ray experiment of Ref.~\onlinecite{chang}.
In addition, Hall effect measurements \cite{leboeuf_hall} indicate a reconstruction of the Fermi surface
that is attributed to the formation of CDW. The transition temperatures $T_{\mathrm{CDW}}$ of these two experiments agree
with each other. Evidence for a transition below~$T^*$ and related to CDW has also been
found recently in a Raman scattering study. \cite{letacon_dec13} The dual effect of the two modulations (QDW and CDW) on the two species
of atoms in the CuO plane is a characteristic of our theory and might be tested
via resonant soft X-ray scattering.

Very recent STM and resonant elastic X-ray experiments \cite{comin,yazdani_dec13} on BSCCO confirm the CDW wave vectors' orientation along the bonds but
indicate that they connect hotspots rather than antinodes. In our
model, we expect CDW to set in at wave vectors as soon as the QDW gap
is small. In realistic systems, this may indeed happen already not
very far from the hotspots, possibly enhanced by reconstruction of the
Fermi surface. Details behind this physics are clearly beyond the range of our ``minimal model'' and left for a separate study.

The emergence of various gaps in $\mathbf{k}$-space around the Fermi surface has been
reported in Raman scattering on Bi-2212 and Hg-1201 compounds. \cite{sacuto}
It was demonstrated that in overdoped samples the superconducting gap spreads
all over the Fermi surface. In contrast, in underdoped samples the coherent Cooper pairs are
observed mostly near the nodes, whereas the gap at the antinodes is mainly of a
non-superconducting origin. This effect can naturally be explained within
our picture because the hotspots move to nodes with decreasing the doping
and the superconducting gap at the antinodes should decrease. At the same
time, the CDW gap grows at the antinodes thus \textquotedblleft pushing
away" the Cooper pairs.\bigskip

We note that after our work has been completed and distributed as a preprint on arXiv, a work discussing the issue of the
rotation of the charge order wave vector by $45^{\circ}$ has appeared. \cite{chubukov2014} A solution of mean-field equations for a new CDW suggested in the latter work, although very interesting, is not stable against formation of SC/QDW order of Ref. \onlinecite{emp} below its transition temperature $T^*$. As a result, new preemptive states predicted in 
Ref.~\onlinecite{chubukov2014} may be possible only in the vicinity of $T^*$.

\section{Conclusion} 

Extending the analysis of the spin-fermion model for the two-dimensional antiferromagnetic QCP
to the antinodal regions, we find below the pseudogap temperature~$T^*$ another transition
to a bidirectional CDW induced at the zone edge by superconducting fluctuations. The physics behind
this transition is determined by pseudogap physics emerging at the hotspots. Our theory
thus shows how a complexity of offspring phases arises out of the single QCP.
The results enable us to address recently observed charge order features in the phase diagram of the high-$T_c$ cuprates.

\acknowledgments

K.B.E. acknowledges support by the Chaire Blaise Pascal award of the R\'{e}gion \^{I}le-de-France. H.M. acknowledges the Yale Prize Postdoctoral Fellowship.
Financial support (K.B.E., H.M., and M.E.) by SFB/TR12 of DFG is gratefully appreciated.

%
%

\end{document}